\documentclass{bmcart}
\usepackage{color}

\usepackage{amsthm,amsmath}
\usepackage[utf8]{inputenc} 
\usepackage{graphicx,booktabs,multirow}
\usepackage{cite}
\usepackage[noend]{algpseudocode}
\usepackage{algorithmicx,algorithm}
\usepackage{subfigure}   

\startlocaldefs
\endlocaldefs

\begin{document}
\begin{frontmatter}

\begin{fmbox}
\dochead{Research}


\title{Audio Tampering Detection Based on Shallow and Deep Feature Representation Learning}

\author[
addressref={aff1},
corref={aff1},
email={zfwang@ccnu.edu.cn}
]{\inits{Z}\fnm{Zhifeng} \snm{Wang}}
\author[
addressref={aff2},                   
email={857861112@qq.com}   
]{\inits{Y}\fnm{Yao} \snm{Yang}}
\author[
addressref={aff2},                   
corref={aff2},
email={cyzeng@hbut.edu.cn}   
]{\inits{C}\fnm{Chunyan} \snm{Zeng}}
\author[
addressref={aff2},                   
email={528444265@qq.com}   
]{\inits{S}\fnm{Shuai} \snm{Kong}}
\author[
addressref={aff2},                   
email={1069643975@qq.com}   
]{\inits{S}\fnm{Shixiong} \snm{Feng}}
\author[
addressref={aff2},                   
email={nzhao@mail.hbut.edu.cn}   
]{\inits{N}\fnm{Nan} \snm{Zhao}}

\address[id=aff1]{%
	\orgname{Department of Digital Media Technology, Central China Normal University},
	\street{Luoyu Road 152},
	\postcode{430079}
	\city{Wuhan},
	\cny{China}
}
\address[id=aff2]{
	\orgname{Hubei Key Laboratory for High-efficiency Utilization of Solar Energy and Operation Control of Energy Storage System, Hubei University of Technology}, 
	\street{Nanli Road 28},                     %
	\postcode{430068}                                
	\city{Wuhan},                              
	\cny{China}                                    
}



\end{fmbox}


\begin{abstractbox}

\begin{abstract} 

Digital audio tampering detection can be used to verify the authenticity of digital audio. However, most current methods use standard electronic network frequency (ENF) databases for visual comparison analysis of ENF continuity of digital audio or perform feature extraction for classification by machine learning methods. ENF databases are usually tricky to obtain, visual methods have weak feature representation, and machine learning methods have more information loss in features, resulting in low detection accuracy. This paper proposes a fusion method of shallow and deep features to fully use ENF information by exploiting the complementary nature of features at different levels to more accurately describe the changes in inconsistency produced by tampering operations to raw digital audio. Firstly, the audio signal is band-pass filtered to obtain the ENF component. Then the discrete Fourier transform (DFT) and Hilbert transform are performed to obtain the phase and instantaneous frequency of the ENF component. Secondly, the mean value of the sequence variation is used as the shallow feature; the feature matrix obtained by framing and reshaping of the ENF sequence is used as the input of the convolutional neural network; the characteristics of the fitted coefficients are obtained by curve fitting. Then, the local details of ENF are obtained from the feature matrix by the convolutional neural network, and the global information of ENF is obtained by fitting coefficient features through deep neural network (DNN). The depth features of ENF are composed of ENF global information and local information together. The shallow and deep features are fused using an attention mechanism to give greater weights to features useful for classification and suppress invalid features. Finally, the tampered audio is detected by downscaling and fitting with a DNN containing two fully connected layers, and classification is performed using a Softmax layer. The method achieves 97.03\% accuracy on three classic databases: Carioca 1, Carioca 2, and New Spanish. In addition, we have achieved an accuracy of 88.31\% on the newly constructed database GAUDI-DI. Experimental results show that the proposed method is superior to the state-of-the-art method.
\end{abstract}


\begin{keyword}
\kwd{Electronic Network Frequency}
\kwd{Audio forensics}
\kwd{Deep learning}
\kwd{Feature fusion}
\end{keyword}


\end{abstractbox}
%

\end{frontmatter}



\section{Introduction}
More and more software for digital audio editing has been developed recently, and it has become much easier to edit, tamper and forge digital audio \cite{Zeng2022a}. The audio editing software (such as Adobe Audition, WavePad, and Ocenaudio) makes it easy for ordinary people to delete, insert, copy and paste digital audio tampering, resulting in changes in audio semantics \cite{Wang2022t}. Moreover, with the continuous enhancement of audio editing technology, it is impossible to tell whether the audio has been tampered with by the human ear \cite{Zeng2022d}. However, some edited digital audio may be misused, especially in essential security applications such as courts, politics, or business, which may cause serious consequences \cite{Qamhan2021}. For digital audio that is deliberately or even maliciously tampered with, it is essential to develop efficient digital audio tampering detection methods \cite{Zeng2022c}.

\par
 Digital audio tampering detection methods include active detection and passive detection \cite{Zeng2020,Zeng2021b}. There are two standard technologies in active detetion methods: digital watermark and digital signature of digital audio. These two technologies require a pre-embedding watermark or signature in the audio to be detected. However, most of the audio to be detected is not pre-embedded with these additional information in practice. The passive detection methods are more practical to use the characteristics of digital audio itself to detect tampering without adding any additional information \cite{Hua2021}.
 \par
At present, passive audio tamper detection methods are mainly based on visual contrast analysis of frequency continuity of electronic network based on digital audio and standard ENF database, or feature extraction of ENF signals and classification by machine learning method. ENF database is usually challenging to obtain, and the feature expression of the visualization method is weak. In contrast, the feature information loss of the machine learning method is significant, resulting in low detection accuracy. This paper proposes an audio tamper detection method based on the fusion of shallow and deep features to solve this problem. Firstly, the audio signal is bandpass filtered to obtain the ENF component. Then the phase and instantaneous frequency of the ENF component are obtained by DFT and Hilbert transform. Second, the ENF phase and instantaneous frequency are processed in three ways. The mean value of sequence variation is taken as the shallow feature. The feature matrix obtained by framing and reshaping of the ENF sequence is the input to the convolutional neural network. Curve fitting was carried out to obtain the characteristics of fitting coefficients. The problem of the unequal length of ENF features is proposed to be solved by framing and fitting processing methods to make them suitable for the input of neural networks. Then, in the neural network, the feature matrix is input into the convolutional neural network to obtain the local details of ENF, and the fitting coefficient feature obtains the global information of ENF through DNN. The global information and local details together constitute the deep features of ENF. The characteristics of the ENF phase and frequency features are fully considered to obtain deep features containing both global and local information about ENF. The attention mechanism fuses the shallow and deep features. The fusion of shallow and deep features exploits the complementary nature of features at different levels to more accurately describe the changes in inconsistency produced by tampering operations to natural digital audio. Finally, the DNN classifier with two fully connected layers was used to fit, and Softmax was used to classify and detect tampered audio. 
The main contributions of this paper are as follows:
\begin{itemize}
	\item[1).] We propose a novel shallow and deep feature fusion based framework for digital audio tampering detection by automaticly analysing the continuity of ENF.
	With using deep learning methods to learn features for tampering detection automatically, the algorithm has a higher degree of automation than threshold and visualization methods. In addition, the proposed framework achieves state-of-the-art performance on three publicly available dataset Carioca 1, Carioca 2, and New Spanish.
	\item[2).] On the one hand, through the fusion of shallow and deep features, it will acquire the complementarity of different features, which is a more comprehensive description of audio ENF features and can be used to improve algorithm robustness and model generalization capabilities. On the other hand, the local and global information are obtained from the audio ENF through automatic learning to reduce information loss and further improve detection accuracy.
	\item[3).] The Attention mechanism is used to fuse phase and frequency features to obtain useful detailed information for tampering detection and classification task from audio through automatic learning to improve classification accuracy and model generalization ability.
\end{itemize}

\par
The rest of this paper is organised as follows. Section \ref{related} presents the relevant existing works in the literature. Section \ref{framework} describes the audio tampering detection framework. Section \ref{method} presents the proposed audio tampering detection method based on shallow and deep feature fusion. The results of experiments and analyses are shown in Section \ref{exp}. Lastly, we come to a conclusion and list some future work in Section \ref{conclusion}.  

\section{Related work} \label{related}
Digital audio passive forensics realizes tampering detection by extracting and analyzing audio features. These features can be divided into traditional shallow features and deep features generated by deep neural networks.

\subsection{Detection methods based on shallow features}
The features contained in digital audio are divided into three categories, which are 1). Environment and device characteristics; 2). Time-domain and frequency domain characteristics; 3). Electronic network frequency characteristics.
\par
\emph{1). Environment and device characteristics in audio:} Digital audio is obtained by recording equipment in a particular environment, which will lead to the existence of some equipment and environment information in the audio \cite{Wang2020h,Zeng2018,Wang2015a,Wang2015b}. An audio file are regarded as edited one when there are different background information involved in the audio \cite{2021zenga}. Malik \cite{Malik2013} carries out endpoint detection of speech signals, extracts the attenuating signal part at the end, and uses statistical methods to model and estimate the reverberation and background noise in the attenuating signal, which is used to classify different signals. In \cite{Zhao2013}, the method is tested and improved on this basis, and the robustness of the original method against MP3 compression is improved. The device information in the audio can be analyzed to determine whether the audio has been edited \cite{Zeng2021a,Wang2021m}. Cuccovillo et al. \cite{luca2013} analyzes the microphones of recorded audio to detect the presence of multiple microphones in single audio for tampering detection. When recording audio, the surrounding environment mainly contains background noise, which can be used for audio tampering detection by analyzing background noise. In \cite{2018DetectingA}, according to the significant differences in the audio background noise levels of different recording environments, the similarity of each syllable's background noise variance is compared to judge whether there is a heterogeneous splicing tampering operation in the audio.

\par
\emph{2). Time-domain and frequency domain characteristics of audio:} When editing digital audio, some features of the audio will be affected, resulting in abnormal changes in features, which will make the discontinuity or correlation between adjacent frames weakened. Audio tampering detection can be realized by analyzing audio time-domain and frequency domain characteristics \cite{Zakariah2017}. Time-domain features are used for tampering detection as follows: Yan et al. \cite{Yan2019} detects the smoothing processing after audio tampering through the local variance of differential signals.Yan et al. \cite{Yan2019a} took pitch sequence and formant sequence as the features of voiced fragments, and realized copy-move tampering detection and location by comparing their similarity. The use of frequency domain features for tampering detection includes: Chen et al. \cite{Chen2014} used wavelet packet singularity analysis to detect the insertion, deletion, replacement, and concatenation operations according to the singularity points generated by the weakened signal correlation caused by audio tampering. In \cite{Lin2017}, Lin et al. used the short time Fourier transform (STFT) to reconstruct the spectral phase to offset the influence of noise, and proposes a feature based on the spectral phase residual and spectral phase correlation between two adjacent clear segments, so as to realize tampering detection and location at the high noise level. Xie et al. \cite{2018CopyM} combined the four characteristics of the Gammatone feature, Mel-frequency cepstral coefficients (MFCCs) feature, pitch feature, and DFT coefficients and adopted the decision tree method to conduct copy-move tampering detection.

\par
\emph{3). Electronic network frequency in audio:} 
ENF is widely used in audio forensics \cite{Wang2018}. ENF is the transmission frequency with a nominal value of 50 or 60hz in the power grid. When recording audio in the electrical activity area, the ENF signal will be embedded into the audio \cite{Hajj-Ahmad2019}. The fluctuation of ENF in a specific area is stable and unique within a certain period \cite{2013SeeingE}, so ENF can be used to detect audio tampering \cite{2020huaa,2021huaa}. Two existing methods for audio tampering detection using ENF include database comparison and consistency analysis. Audio tampering detection can be carried out by comparing ENF in audio with the ENF database. Hua et al. \cite{Hua2016} detects insertion, deletion, and stitching operations through the Absolute-Error-Map between the ENF signal in audio and the database. However, it is difficult to obtain the ENF database, and many studies have used ENF discontinuity to detect audio. Most tampering operations cause the ENF to change suddenly at the tampering point. Esquef et al. \cite{2014EditD} uses Hilbert transform to calculate the instantaneous frequency of ENF and proposes TPSW (two-pass Split Window) method to estimate the change degree ENF background to achieve tampering detection. Rodriguez et al. \cite{2010AudioA, Nicolalde2009} detects audio tampering by extracting ENF signals and detecting ENF phase changes' consistency.

\subsection{Detection methods based on deep features}
With the development of deep learning and artificial intelligence technology \cite{Zeng2022,Zeng2022b,Wang2022ah,Wang2022ag,Wang2022ac,Lyu2022}, some scholars also use deep learning methods to deal with audio forensics tasks. Deep learning-based methods perform audio tampering detection tasks by training a deep neural network model with data set in advance. Using a large amount of data to train the model can reduce the practical problems caused by artificially setting the threshold. The deep learning method supports higher-dimensional input features \cite{Zeng2021c,Wang2021,Wang2017,Wang2015a,Zeng2020x}. Combining multiple parameters in the deep neural network can better fit the audio features, learn the difference between original audio and tampered audio, and make the detection more accurate and more robust.
\par
Tamper detection based on deep learning methods can be divided into three subcategories.
\emph{1). Frequency domain features are used to identify audio post-processing operations.} Wang et al. \cite{Wang2018a} used the features of audio after STFT transformation as the input of the convolutional neural network (CNN) to identify the post-processing operation of audio pitch transformation. \emph{2). ENF is applied for audio recapture detection.} Lin et al. \cite{2016AudioR} takes ENF spectrogram as the convolutional neural network input for audio recapture detection. \emph{3). Use the spectrogram to detect insertion and tampering in audio.} Jadhav et al. \cite{2019AudioS} directly input the audio spectrum map into the convolutional neural network to detect the audio insertion tampering.

\begin{figure}[ht]
    \includegraphics[scale=0.18]{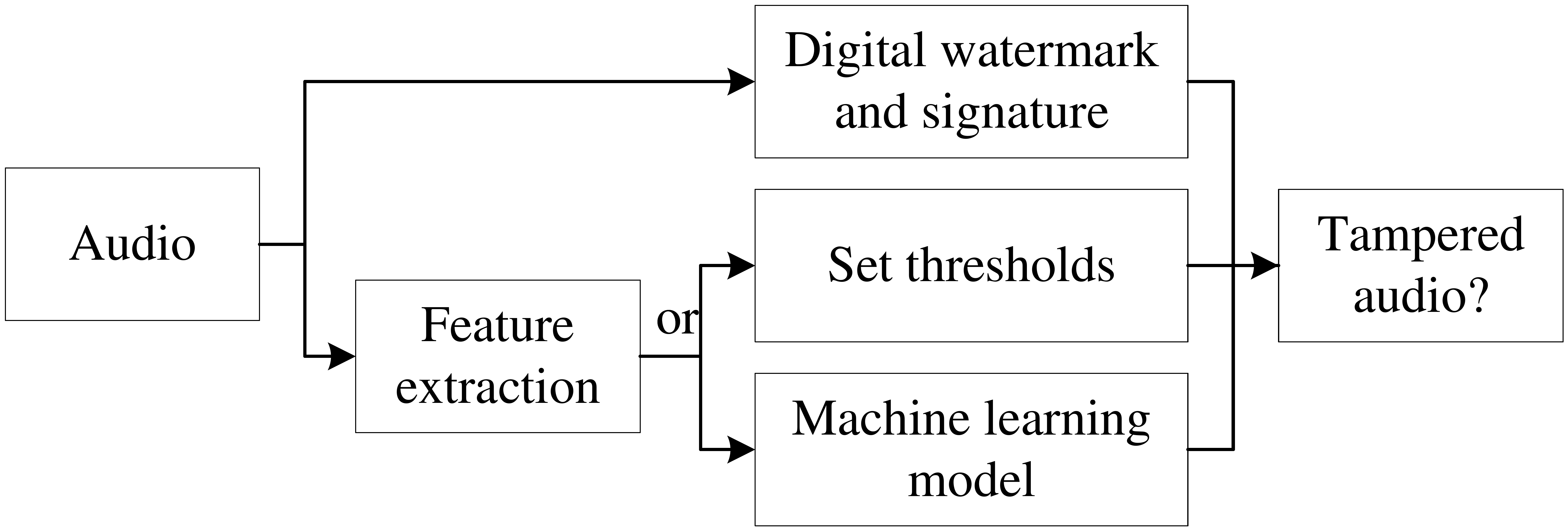}
    \caption{Audio tampering detection framework.}
    \label{fig:1}
\end{figure}

\section{Audio tampering detection framework} \label{framework}
The audio tampering detection framework is shown in Fig.\ref{fig:1}. The tampering detection methods can be classified into active detection methods and passive detection methods. The active detection method is to embed a digital watermark, and signature in the audio when the audio is generated. When the audio is edited and tampered with, the watermark information embedded in the audio in advance will change so that the edited audio can be accurately distinguished. However, there are often no such watermarks in audio. The passive detection method uses the audio content itself as a feature, detects tampering in these features through a threshold, or trains a model through machine learning and other methods to perform tampering detection.
\par
In the audio tampering detection task, the audio signal can be formulated by
\begin{equation}
y\left( n \right) = s\left( n \right) + v\left( n \right) + f\left( n \right)
\label{eq1_}
\end{equation}
Where $s\left( n \right)$ represents speech content, $v\left( n \right)$ represents background noise, and $f\left( n \right)$ represents ENF. In traditional digital audio tampering detection, one of the speech content, background noise or ENF in audio, is usually extracted and analyzed. The audio is windowed and divided into frames. Extract the time domain or frequency domain features of each frame of audio, such as pitch, reverberation, background noise, MFCC, ENF, and other time and frequency domain features. Then set the corresponding threshold to detect the abrupt changes between frames or detect it through the support vector machine (SVM) \cite{Wang2018}.

\section{Audio tampering detection based on shallow and deep feature fusion} \label{method}
The audio tampering detection method based on the fusion of shallow and deep features proposed in this paper consists of three parts (Fig.\ref{fig:2}):
\begin{figure}[ht]
    \includegraphics[scale=0.8]{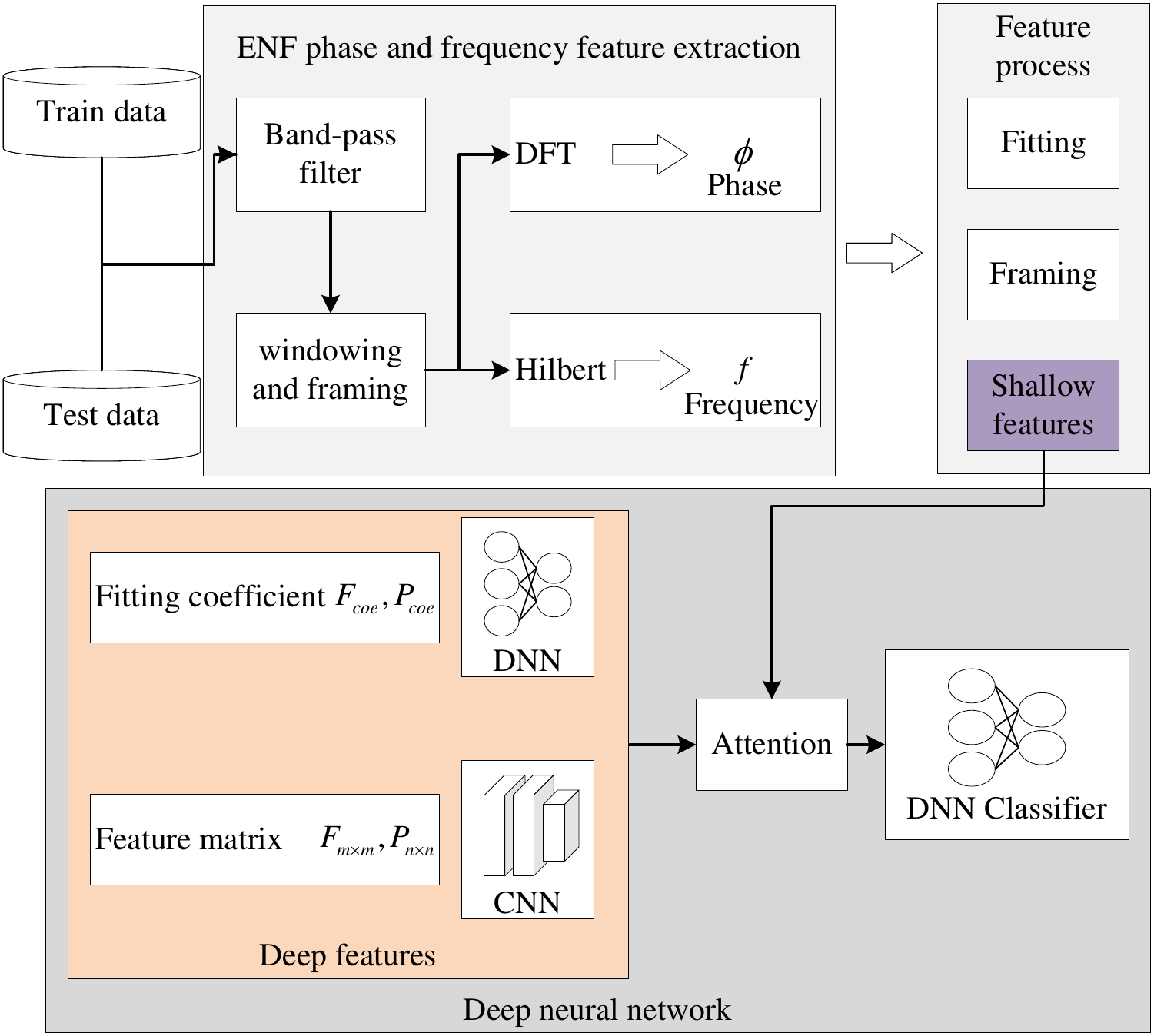}
    \caption{Shallow and deep feature fusion method}
    \label{fig:2}
\end{figure}

\begin{itemize}

\item[1).] Phase and frequency feature extraction: First, down-sampling and band-pass are employed to filter the audio to obtain the ENF component. Then windowing and framing process are implemented on the ENF components. Finally, DFT is used to obtain the phase feature, and Hilbert transform is applied to obtain the instantaneous frequency.

\item[2).] Feature process: In this part, the average of ENF phase and frequency variations is calculated as the shallow features. Meanwhile, the feature matrix is obtained by framing and reshaping operations on the audio (see the section \ref{tefea} for details). The feature matrix will be used as the convolutional neural network's input to obtain more local information. The fit coefficients are obtained by fitting the phase to the instantaneous frequency through Sums of Sines functions \cite{Wang2018}, and the fit coefficients are the input to the DNN to give some global information compensation to the deeper features.

\item[3).] Deep neural network:In this part, the feature matrix and fitting coefficients are input to the neural network, and the output is stitched to obtain the deep features that contain both global and local information. Finally, the deep, deep phase, and instantaneous frequency features are fused with features using the Attention mechanism to give different weights to each feature vector value to achieve feature selection. Finally, a DNN classifier is proposed to classify the tampered audio with the authentic audio.
\end{itemize}
\par
The specific details will be introduced later in this section.

\subsection{ENF phase and frequency feature extraction}

According to the method in literature \cite{2010AudioA,2014EditD}, to obtain the phase and instantaneous frequency of ENF, we performed subsampling and bandpass filtering on the audio. Firstly, the subsampling frequency is set as 1000Hz and 1200Hz according to the ENF nominal frequency of 50 or 60Hz. The purpose of this is to ensure the accuracy of ENF while reducing the amount of calculation. Then, after subsampling, bandpass filtering is carried out to obtain ENF components in the audio. We use a linear zero-phase FIR filter of order 10000 to carry out narrowband filtering. The center frequency is ENF standard (50Hz or 60Hz), the bandwidth is 0.6Hz, the passband ripple is 0.5dB, and the stopband attenuation is 100dB. Finally, we can obtain the phase and instantaneous frequency of ENF through DFT and Hilbert transformation.

\subsubsection{DFT transformation gets the phase}
The phase of ENF is obtained by discrete Fourier transform, and the phase of $DF{T^0}$ and $DF{T^{\rm{1}}}$ is calculated. $DF{T^k}$ represents the $k$ derivative of the $DFT$ transform of a signal, and $DF{T^0}$ represents the conventional $DFT$ transform \cite{2010AudioA}.

\par
First, the approximate first derivative ${x'_{ENFC}}\left[ n \right]$ of ENF signal ${X_{ENFC}}\left[ n \right]$ at point $n$ is calculated 
\begin{equation}
{x'_{ENFC}}\left[ n \right] = {f_d}\left( {{X_{ENFC}}\left[ n \right] - {X_{ENFC}}\left[ {n - 1} \right]} \right) 
\label{eq1}
\end{equation}
Where ${f_d}\left(  *  \right)$ represents the approximate derivative operation, and ${X_{ENFC}}\left[ n \right]$ represents the n-th point of the ENF component.
\par
Then, Hanning window $w\left( n \right)$ was used to frame and window ${x'_{ENFC}}\left[ n \right]$. The frame length was 10 standard ENF frequency cycles ($\frac{{10}}{{50}}$ or $\frac{{10}}{{60}}$), and the frame was moved to 1 standard ENF frequency cycle ($\frac{1}{{50}}$ or $\frac{1}{{60}}$).
\begin{equation}
{x'_N}\left[ n \right] = {x'_{ENFC}}\left[ n \right]w\left( n \right)
\label{eq2}
\end{equation}
Where ${x'_N}\left[ n \right]$ represents the ENF signal after window addition, and $w\left( n \right)$ represents the Hanning window.
\par
To obtain the phase ${\phi _{DF{T^0}}}$ of ENF and the phase ${\phi _{DF{T^1}}}$ of the first derivative of ENF, n-point DFT should be executed for each frame signal ${x'_N}\left[ n \right]$ and ${X_{ENFC}}\left[ n \right]$ respectively to obtain $X'\left( k \right)$ and $X\left( k \right)$. Estimated frequency ${f_{DF{T^1}}}$ based on the integer index ${k_{peak}}$ of $\left| {X'\left( k \right)} \right|$ peak points
\begin{equation}
{f_{DF{T^1}}} = \frac{1}{{2\pi }}\frac{{DF{T^1}\left[ {{k_{peak}}} \right]}}{{DF{T^0}\left[ {{k_{peak}}} \right]}}
\label{eq3}
\end{equation}
Where, $DF{T^0}\left[ {{k_{peak}}} \right] = X\left( {{k_{peak}}} \right)$, $DF{T^1}\left[ {{k_{peak}}} \right] = F\left( {{k_{peak}}} \right)\left| {X'\left( {{k_{peak}}} \right)} \right|$ and $F\left( {{k_{peak}}} \right)$ are scale coefficients.
\begin{equation}
F\left( k \right) = \frac{{\pi k}}{{{N_{DFT}}\sin \left( {\frac{{\pi k}}{{{N_{DFT}}}}} \right)}}
\label{eq4}
\end{equation}
Where ${N_{DFT}}$ represents the number of discrete Fourier transform points, and $k$ is the index of peak point.
\par
Now the ENF phase ${\phi _{DF{T^0}}}$ of the conventional DFT transformation can be calculated, ${\phi _{DF{T^0}}} = \arg \left[ {X\left( {{k_{peak}}} \right)} \right]$. Through Equation \eqref{eq5}, ${\phi _{DF{T^1}}}$ \cite{2010AudioA} can be calculated.
\begin{equation}
\left\{ \begin{array}{l}
{\phi _{DF{T^1}}} = \arctan \left\{ {\frac{{\tan \left( \theta  \right)\left[ {1 - \cos \left( {{\omega _0}} \right)} \right] + \sin \left( {{\omega _0}} \right)}}{{1 - \cos \left( {{\omega _0}} \right) - \tan \left( \theta  \right)\sin \left( {{\omega _0}} \right)}}} \right\}\\
\theta  \approx \left( {{k_{DF{T^1}}} - {k_{low}}} \right)\frac{{{\theta _{high}} - {\theta _{low}}}}{{{k_{high}} - {k_{low}}}} + {\theta _{low}}
\end{array} \right.
\label{eq5}
\end{equation}
Where, ${\omega _0} \approx 2\pi {f_{DF{T^1}}}/{f_d}$, ${f_d}$ are heavy sampling frequency, ${k_{DF{T^1}}} = {f_{DF{T^1}}}{N_{DFT}}/{f_d}$, ${k_{low}} = floor\left[ {{k_{DF{T^1}}}} \right]$, ${k_{high}} = ceil\left[ {{k_{DF{T^1}}}} \right]$, $floor\left[ a \right]$ is the maximum integer less than $a$, and $ceil\left[ b \right]$ is the minimum integer greater than $b$. Since the calculated ${\phi _{DF{T^1}}}$ has two possible values, ${\phi _{DF{T^0}}}$ is used as a reference, and the value closest to ${\phi _{DF{T^0}}}$ in ${\phi _{DF{T^1}}}$ is selected as the final ${\phi _{DF{T^1}}}$.

\subsubsection{The Hilbert transform captures the instantaneous frequency}
Hilbert transformation \cite{2014EditD} was performed on the filtered ENF signal ${X_{ENFC}}\left[ n \right]$ to obtain the ENF instantaneous frequency $f\left[ n \right]$. So first we get the analytic function of ${X_{ENFC}}\left[ n \right]$
\begin{equation}
{x^{\left( a \right)}}_{ENFC}\left[ n \right] = {X_{ENFC}}\left[ n \right] + i*H\left\{ {{X_{ENFC}}\left[ n \right]} \right\}
\label{eq6}
\end{equation}
Where $H\left\{  *  \right\}$ stands for Hilbert transformation, $i = \sqrt { - 1} $. Instantaneous frequency $f\left[ n \right]$ is the rate of change of $H\left\{ {{X_{ENFC}}\left[ n \right]} \right\}$ phase Angle.
\par
The parasitic oscillation generated by the numerical approximation during the Hilbert transformation needs to be removed after the instantaneous frequency $f\left[ n \right]$ obtained. The fifth-order elliptic IIR filter was used to carry out the low-pass filter on $f\left[ n \right]$ to remove oscillation. The filter's central frequency is ENF standard frequency, the bandwidth is 20Hz, the passband ripple is 0.5dB, and the stopband attenuation is 64dB. Due to the boundary effect of frequency estimation, the head and tail of $f\left[ n \right]$  are removed for about 1s. Finally, ${f_{hil}}$ of instantaneous frequency estimation of ENF component is obtained.

\subsection{Shallow feature acquisition and deep feature preparation}
We use the average of ENF phase and instantaneous frequency changes as shallow features. To obtain the deep features, we use a convolutional neural network better to learn the details of ENF phases and instantaneous frequencies. We frame, Reshape and fit the ENF phase and frequency to get the input to the neural network and feed it to the neural network to obtain the depth features for the training phase of the network.

\subsubsection{Acquire shallow features}
The estimated phase ${\phi _{DF{T^0}}}$, ${\phi _{DF{T^1}}}$ and Hilbert instantaneous frequency ${f_{hil}}$ are put into Equation \ref{eq9} to obtain the statistical feature ${F_{01f}} = \left[ {{F_0},{F_1},{F_f}} \right]$, which reflects the abrupt transition of ENF phase and instantaneous frequency [19].
\begin{equation}
\left\{ \begin{array}{l}
{F_{0,1}} = 100\log \left\{ {\frac{1}{{{N_{Block}} - 1}}\sum\limits_{{n_b} = 2}^{{N_{Block}}} {{{\left[ {\hat \phi '\left( {{n_b}} \right) - {m_{\hat \phi '}}} \right]}^2}} } \right\}\\
{F_f} = 100\log \left\{ {\frac{1}{{len - 1}}\sum\limits_{n = 2}^{len} {{{\left[ {f'\left( n \right) - {m_{f'}}} \right]}^2}} } \right\}
\end{array} \right.
\label{eq9}
\end{equation}
Where $\hat \phi '\left( {{n_b}} \right) = \hat \phi \left( {{n_b}} \right) - \hat \phi \left( {{n_b} - 1} \right)$, $2 \le {n_b} \le {N_{Block}}$. $\hat \phi \left( {{n_b}} \right)$ is the estimated phase of the corresponding ${n_b}$ frame. ${m_{\hat \phi '}}$ represents the average value of $\hat \phi '\left( {{n_b}} \right)$ from ${n_b} = 2$ to ${N_{Block}}$. $len = length({X_{ENFC}}\left[ n \right])$, $f'\left( n \right) = f\left( n \right) - f\left( {n - 1} \right)$. $f\left( n \right)$ is the instantaneous frequency of the n-th sampling point, and ${m_{f'}}$ represents the average value of $f'\left( n \right)$ from $n = 2$ to $len$.

\subsubsection{Obtaining the input of deep features ${F_{m \times m}}$,${P_{n \times n}}$} \label{tefea}
The deep features proposed in this paper consist of two parts, firstly, the local detail information obtained by the feature matrix ${F_{m \times m}}$ and ${P_{n \times n}}$ through the convolutional neural network, obtained by framing and reshaping operations. The second is the global information obtained by fitting coefficients through DNN. Finally, the global information is stitched with detailed information to obtain deep features.
\par
To reduce information loss, we acquire the deep features by convolutional neural networks. Therefore, we designed a framing approach for obtaining the input of the convolutional neural network so that the audio ENF phase or frequency of unequal lengths through the dataset becomes a matrix of $m \times m$. Where m is the frame length (the audio determines the frame length with the longest duration in the data), and each row in the matrix is one frame, and the frame shift s of each audio is computed adaptively. The detailed steps are listed in the following Algorithm \ref{suanfa}.

\begin{algorithm}[!h]
\caption{Obtaining the input of deep features ${F_{m \times m}}$,${P_{n \times n}}$} 

\begin{algorithmic}[1]
\State dataset:The audio length is about 9-35s 
\For{All audio data} 
    \State Get the maximum length of audio
\EndFor
\State DFT and Hilbert transform for the longest duration audio
\State Get the maximum ${\phi}$ length $len{\left( \phi  \right)_{\max }}$ and ${f'}$ length $len{\left( f \right)_{\max }}$
\State Calculates the frame length $m\left( n \right) = ceil\left( {\sqrt {{X_{\max }}} } \right),{X_{\max }} = len{(\phi )_{\max }},len{\left( f \right)_{\max }}$
\For{All audio data} 
    \State DFT and Hilbert transform 
    \State Calculate the overlap and divide the frame. ${\rm{overlap}} = m\left( n \right) - ceil\left( {\frac{{X - m\left( n \right)}}{{m\left( n \right) - 1}}} \right)$
    Reshape into feature matrix ${F_{m \times m}}$,${P_{n \times n}}$
\EndFor

\State \Return ${F_{m \times m}}$,${P_{n \times n}}$
\end{algorithmic}
\label{suanfa}
\end{algorithm}

\subsubsection{Curve fitting for fitting coefficient}
We performed a reshape operation when obtaining the feature matrix of the convolutional neural network input, which may result in the loss of global information of the sequence, so we fit the ENF phase and frequency sequences and used the fit coefficients as compensation for the global information. The ENF phase and instantaneous frequency are curve-fitted to extract the fit coefficients containing the global information. We use the Matlab fitting toolbox to extract the fitting coefficients using six Sum of Sines functions to fit the phase, and frequency features ${F_{{\rm{coe}}}},{P_{coe}} = \left[ {{a_1},{b_1},{c_1}, \cdots ,{a_j},{b_j},{c_j}} \right]\left( {0 < j \le 6} \right)$. the Sum of Sines functions are

\begin{equation}
y = \sum\limits_{j = 1}^6 {{a_j}\sin \left( {{b_j}x + {c_j}} \right)}
\label{eq7}
\end{equation}

\begin{figure}[!ht]
    \centering
    \includegraphics[scale=0.8]{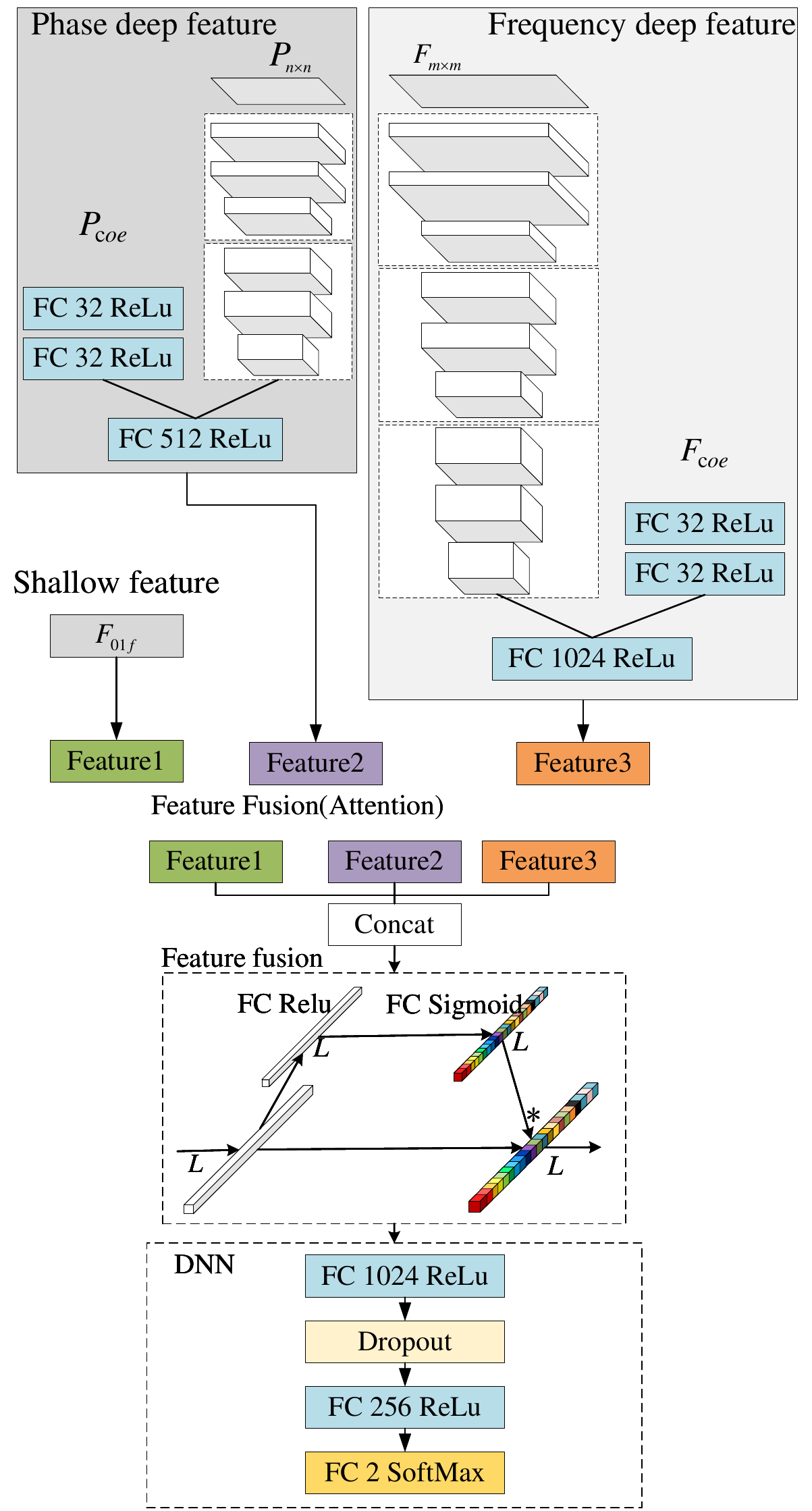}
    \caption{Shallow and deep feature fusion network}
    \label{fig:3}
\end{figure}

\subsection{Shallow and deep feature fusion network}

There is information loss by only going through shallow features, resulting in the inability to obtain higher detection accuracy and model generalization. The duration of each detected audio is different, so the obtained phase feature length and frequency feature length are also different. As shown in Fig.\ref{fig:3}, in the tampering detection method based on the fusion of shallow and deep features proposed in this paper, the phase and instantaneous frequency features of the ENF are first processed to make them suitable for automatic learning of the neural network and reduce information loss. Then the depth features of ENF are obtained by the neural network to understand better the difference between tampered audio and real audio by automatic learning. Then feature fusion is performed using Attention, and finally, the detection results are output.

\subsubsection{Neural networks of deep features}
As shown in Fig.\ref{fig:3}, the shallow feature ${F_{123}}$, which are extracted through the framing and the sum of sines fitting, reflects the sudden change of ENF phase and frequency, but its statistical feature is only a single value, and detailed information about ENF phase and frequency will be lost. When it is only used for audio tampering detection, it may cause misjudgment due to the insignificant fluctuation of ENF in the tampered area, or the interference of low-frequency noise on ENF. In order to reduce the occurrence of this situation, we use the convolutional neural network to obtain ENF detailed information as deep features and use the attention mechanism to combine deep features with shallow features to reduce misjudgments and improve the generalization ability of the model.
\par
The deep features proposed in this paper are obtained from the fitting coefficients and the feature matrix. The fitting coefficients are passed through two fully connected layers with 32 neurons to obtain the ENF phase and global frequency information. A convolutional neural network extracts the phase and frequency feature matrices to obtain detailed information about the ENF phase and frequency. The size of the phase feature matrix ${P_{n \times n}}$ is different from that of the frequency feature matrix ${F_{m \times m}}$. As the size of the feature matrix is $n \times n$,$m \times m$, the frame length is determined by the longest audio in the audio data, and the longest duration of the digital audio that this network can detect is 35s. Since the longest audio in our data set is 35s, the length of the phase and instantaneous frequency sequences obtained by DFT and Hilbert transform are 2055 and 37281, so our frame length in the deep feature is set to 46,194 by the steps in \ref{suanfa}. The number of convolution blocks for phase features is 2, and for instantaneous frequency, convolution blocks are 3. When the longest length of the audio to be measured increases or decreases, the number of convolution blocks should be increased or decreased as appropriate.
\par
We use two convolution blocks to extract features from the phase feature matrix ${P_{n \times n}}$ and three convolution blocks to extract features from the frequency feature matrix ${F_{m \times m}}$. Each convolutional block consists of two identical convolutional layers with one pooling layer (the number of filters for the three convolutional blocks is 32,64,128. The convolutional kernel size is 3*3 and the step is 1. The Maxpooling layer pool size is 3). detailed information of the ENF phase and frequency sequence can be obtained by using the local sensing property of the convolutional neural network. The pooling layer is used for dimensionality reduction to reduce the number of parameters, avoid overfitting, and improve the model's fault tolerance and generalization ability. Also, because the convolutional neural network has fewer parameters, it can obtain better classification results with less training time.
\par
Frequency fitting coefficient ${F_{coe}}$, two fully connected layers were used to fit its characteristics (the number of neurons was 32, 32, and the activation function was Relu). The output of the convolution block is dimensioned through a layer of fully connected with 1024 neurons, then splicing with the fitting coefficient features after DNN fitting. Finally, the deep feature is obtained through the fully connected layer of 1024 neurons. The deep feature contains both the global information of the fitting coefficient and the local information obtained by the convolutional neural network.
 
\subsubsection{The attention mechanism of feature fusion}
We use the attention mechanism \cite{vaswaniAttentionAllYou2017} to fuse shallow and deep features. In the feature fusion part(As shown in Fig.\ref{fig:3}), firstly, we concatenate the shallow and deep features of phase and frequency to obtain the input of length L. Then, to get the weight of each feature, we will input the fully connected layer through the two activation functions for ReLU and Sigmoid. We use the ReLU activation function to enhance the nonlinearity and obtain the weight through Sigmoid. Finally, the input features are multiplied by the weights.
\par

The attention fusion mechanism used in this paper uses the sigmoid activation function instead of softmax to obtain the weights because the primary purpose of the attention mechanism used in this paper is to suppress invalid features, not to find the optimal features. There is no need for each feature value in the shallow and deep layers to compete for weights. This is because the primary purpose of the attention mechanism used in this paper is to suppress the invalid features, not to find the optimal features.The attention fusion mechanism in this paper can automatically learn to give different weights to each feature value of the shallow and deep features. The features significantly impacting the classification result will be given a larger weight. In comparison, the features that do not significantly affect the final classification will be given a smaller weight to improve the detection accuracy and generalization ability.

\subsubsection{DNN classifier}
We use the attention mechanism to fuse shallow and deep features. In the feature fusion part(As shown in Fig.\ref{fig:3}), firstly, we concatenate the shallow and deep features of phase and frequency to obtain the input of length L. Then, to get the weight of each feature, we will input the fully connected layer through the two activation functions for ReLU and Sigmoid. We use the ReLU activation function to enhance the nonlinearity and obtain the weight through Sigmoid. Finally, the input features are multiplied by the weights. Through automatic learning, we give different weights to each value of shallow and deep features. The features that have a significant impact on classification are given greater weights. In comparison, those that are ineffective in classification are given smaller weights to improve detection accuracy.

\section{Experimental results and discussion} \label{exp}
In this section, we will introduce the data set, experimental setup, and experimental results. In order to verify the validity of our work, we designed five groups of experiments to verify our contribution : 1) comparison of the method proposed in this paper with the state-of-the-art methods, 2) validation of the fitting coefficient feature, 3) validation of the feature matrix, 4) validation of the deep feature, and 5) validation of the attention mechanism.
\subsection{Data set and experimental setup}
In order to verify the effect of the proposed model on different data sets and prove the generalization ability of the proposed model, we use two different data sets as experimental data. They are the classic data set composed of three public data sets and the GAUDI-DI data set we collected. The detailed information is shown in the Table \ref{tab:1}.

\begin{table}[htbp]
	\centering
	\caption{Datasets used in the following experiments.}
	\begin{tabular}{lccl}
		\toprule
		Dataset & \multicolumn{1}{l}{Number of original audio} & \multicolumn{1}{l}{Number of edited audio} & Data source \\
		\midrule
		\multirow{3}[2]{*}{Classical} & \multirow{3}[2]{*}{250} & \multirow{3}[2]{*}{250} & Carioca 1 dataset \\
		&       &       & Carioca 2 dataset \\
		&       &       & New Spanish dataset \\
		\midrule
		GAUDI-DI & 251   & 502   & GAUDI dataset \\
		\bottomrule
	\end{tabular}%
	\label{tab:1}%
\end{table}%

In the table \ref{tab:1}, the classic dataset we used contains 500 audios and is a mixture of three public datasets, including Carioca, 1, 2, and New Spanish dataset (from two public Spanish language datasets AHUMADA and GAUDI). In order to verify the generalization ability of the model, we established a GAUDI-DI dataset containing 753 audios, selected 251 original audios of about 20s from the GAUDI dataset, and performed deletion and insertion tampering operations.

\begin{table}[htbp]
	\centering
	\caption{Training set, validation set, and testing set on the same dataset and across-dataset experiments.}
	\begin{tabular}{lccc}
		\toprule
		Dataset & Training set & Validation set & Testing set \\
		\midrule
		Same dateset testing on Classical & 319   & 80    & 101 \\
		Cross dataset testing on GAUDI-DI & 400(Classical) & 100(Classical) & 753(GAUDI-DI) \\
		\bottomrule
	\end{tabular}%
	\label{tab:2}%
\end{table}%

\par
The experimental data are divided as shown in the table \ref{tab:2}. When testing on the classical dataset, we divide the classical dataset into a training set, a validation set, and a test set with 319 audios in the training set, 80 audios in the validation set, and 101 audios in the test set. When testing with the GAUDI-DI dataset, we will use the classical dataset for training and the GAUDI-DI dataset for testing. The training and validation sets are from the classical dataset, with 400 audios in the training set, 100 audios in the validation set, and 753 audios in the test set.
\par
All the experiments in this paper are based on the TensorFlow 2.0 deep learning framework and performed on NVIDIA GeForce GTX 1080Ti. The specific parameters used in the experiment are as follows:
the loss function is binary cross entropy, and the optimizer uses Adam, epochs are 100, the batch size is 32, learning rate decay: initial learning rate is 0.001, Halve every 10 epochs.

\subsection{Comparison of the proposed method with the state-of-the-art methods.}
In this experiment, we compared the proposed method in this paper with four baseline methods to verify the effectiveness of the proposed method. The comparison experiments are performed on the same dataset, where ${F_0}$ features and SVM classifier are applied in \cite{Nicolalde2009}, ${F_1}$ features and SVM classifier are used in \cite{2010AudioA}, ${F_f}$ features and SVM classifier are employed in \cite{2014EditD}, and ${F_0}$, ${F_1}$, and ${F_f}$ features are fused and SVM classifier is also utilized in \cite{Wang2018}. While ${F_0}$ features and ${F_1}$ features are related to phase features, ${F_f}$ features are related to frequencey features, and their extraction details are in Sections 4.1 and 4.2.

\begin{table}[htbp]
	\centering
	\caption{Comparison with the state-of-the-art methods.}
	\begin{tabular}{lrrrr}
		\toprule
		\multicolumn{1}{c}{\multirow{2}[4]{*}{Methods}} & \multicolumn{2}{c}{Classical Dataset} & \multicolumn{2}{c}{GAUDI-DI Dataset} \\
		\cmidrule{2-5}          & \multicolumn{1}{l}{Accuracy} & \multicolumn{1}{l}{F1-Score} & \multicolumn{1}{l}{Accuracy} & \multicolumn{1}{l}{F1-Score} \\
		\midrule
		Nicolalde et al. \cite{Nicolalde2009} & 92.08\% & 92.54\% & 83.53\% & 78.84\% \\
		Nicolalde et al. \cite{2010AudioA} & 94.06\% & 95.44\% & 83.67\% & 78.90\% \\
		Esquef et al. \cite{2014EditD} & 83.17\% & 82.11\% & 79.02\% & 74.18\% \\
		Wang et al. \cite{Wang2018} & 95.05\% & 95.41\% & 84.99\% & 79.14\% \\
		Our method & \textbf{97.03\%} & \textbf{96.91\%} & \textbf{88.31\%} & \textbf{88.17\%} \\
		\bottomrule
	\end{tabular}%
	\label{tab:3}%
\end{table}%

As shown in Table \ref{tab:3}, the accuracy and F1-score of the proposed method on Classical dataset and GAUDI-DI dataset are higher than the four baseline methods. The best performance among the traditional methods is the one using feature fusion in \cite{Wang2018}. Further, the method in this paper improved accuracy by 2\% to 3.3\% and F1-score by 1.5\% to 9\% on both datasets compared to \cite{Wang2018}. This shows that the proposed method obtains the advantage of fused features along with better feature characterization.

It can be seen that all methods perform better on Classical dataset test than on GAUDI-DI datset. The main reason for this is that the test on GAUDI-DI dataset uses cross-dataset detection, which means the training model is trained with data from Classical dataset and tested with data from GAUDI-DI dataset. The main purpose is to perform generalization performance tests. The experimental results show that although the performance of the proposed method is degraded on the cross-dataset test, it still obtains a good performance, which indicates that the proposed method in this paper has a good generalization performance.

Meanwhile, it can be seen from Table \ref{tab:3} that the accuracy of the frequency feature is significantly lower than that of the phase feature in both experiments. This is because the length of the instantaneous frequency sequence obtained by Hilbert transform is about 18 times that of the phase sequence obtained by DFT transform (the frequency length of 35s audio is 37281, and the phase length is 2055). Tampering detection only by the average value of ENF sequence variation has excessive information loss. The proposed method has less information loss of deep features obtained through neural networks, and the fusion of ENF phase and frequency, shallow and deep features improves the detection accuracy and generalization ability.

\subsection{Validation of fitting coefficient features.}
In this section, we conduct experiments to verify the validity of the fitted coefficient feature, which is a key part of performing deep feature representation learning. The fitting coefficients contain global information of the phase and instantaneous frequency series. The reshape operation of the feature matrix in the deep feature leads to the loss of the general information, so the fitting coefficients are used to compensate for the global information of the deep feature. We fit the ENF phase and frequency series by the Sum of Sines function in MATLAB fitting toolbox, and the number of Sum of Sines function is verified in this section, and the classifiers are SVM, random forest, and XGboost. the results are shown in Table \ref{tab:4}.

\begin{table}[htbp]
	\centering
	\caption{Detection accuracy of fitting coefficient features (\%).}
	\begin{tabular}{lllllllll}
		\toprule
		Feature & Classifier & Dataset & 3 Sines & 4 Sines & 5 Sines & 6 Sines & 7 Sines & 8 Sines \\
		\midrule
		& \multirow{2}[2]{*}{SVM} & Classical & 79.21 & 79.21 & 82.18 & 88.12 & 88.12 & \textbf{91.09} \\
		&       & GAUDI-DI & 77.29 & 79.28 & 86.19 & \textbf{86.59} & 86.32 & 83.40 \\
		\cmidrule{2-9}    Proposed & \multirow{2}[2]{*}{RF} & Classical & 87.13 & 85.15 & 83.17 & 86.14 & 87.13 & \textbf{92.08} \\
		feature $f'$  &       & GAUDI-DI & 83.27 & 82.74 & \textbf{86.85} & 86.59 & 85.13 & 86.59 \\
		\cmidrule{2-9}          & \multirow{2}[2]{*}{Xgboost} & Classical & 88.12 & 88.12 & 84.16 & 89.11 & 87.13 & \textbf{90.10} \\
		&       & GAUDI-DI & 76.49 & 80.08 & 81.67 & 80.48 & \textbf{86.32} & 85.92 \\
		\midrule
		& \multirow{2}[2]{*}{SVM} & Classical & \textbf{86.14} & 78.22 & 86.14 & 80.20 & 81.19 & 79.21 \\
		&       & GAUDI-DI & \textbf{77.95} & 75.83 & 73.04 & 74.63 & 73.71 & 74.50 \\
		\cmidrule{2-9}    Proposed & \multirow{2}[2]{*}{RF} & Classical & 90.10 & 89.11 & 90.10 & \textbf{90.10} & 89.11 & 90.10 \\
		feature $\phi$ &       & GAUDI-DI & 79.42 & 79.68 & 80.08 & \textbf{82.74} & 82.60 & 81.14 \\
		\cmidrule{2-9}          & \multirow{2}[2]{*}{Xgboost} & Classical & \textbf{91.09} & 89.11 & 90.10 & 87.13 & 86.14 & 88.12 \\
		&       & GAUDI-DI & 80.61 & 77.29 & 79.55 & 80.48 & 81.01 & \textbf{81.41} \\
		\bottomrule
	\end{tabular}%
	\label{tab:4}%
\end{table}%

\par

Table \ref{tab:4} shows the results of the experiments on both data sets with the fitted coefficients extracted by fitting with 3 to 8 Sum of Sines functions. When the detection accuracy was verified on Classical, the highest detection accuracy of the instantaneous frequency $f'$ was 92.08\% (8 Sum of Sines functions, SVM method). The highest detection accuracy of the phase ${\phi}$ was 91.09\% (3 Sum of Sines functions, XGboost method). The highest detection accuracy of transient frequency F was 86.85\% (5 Sum of Sines functions, Random Forest (RF) method). The highest detection accuracy of phase ${\phi}$ was 82.74\% (6 Sum of Sines functions, Random Forest method) when the generalization ability was verified on GAUDI-DI. Since our fitted coefficient features are used as global information compensation for deeper features, the purpose is to obtain higher model generalization ability. The accuracy of 86.59\% was also obtained with 6 Sum of Sines functions when the generalization ability was verified gaudi. Therefore, the fitted narrative of the 6 Sum of Sines function selected in this paper is used to compensate for the global information of the deep features.

\par

The fitted coefficient feature has low dimensionality, less computation, and better detection accuracy. Furthermore, it can reduce the ENF phase and instantaneous frequency sequences of different lengths to the same dimension, which is convenient for automated detection. Therefore, the ENF phase and instantaneous frequency sequences can be downscaled by using the fitting coefficient feature, and the global information of the ENF phase and instantaneous frequency can be obtained with less computation.

\subsection{Validation of feature matrix ${F_{m \times m}}$,${P_{n \times n}}$}

In this section, the feature matrix ${F_{m \times m}}$,${P_{n \times n}}$ is obtained in section \ref{tefea}. is validated. The model we used is the DNN classifier added after the last convolution block \ref{fig:3}. The experimental results are shown in Table \ref{tab:5}.

\begin{table}[htbp]
	\centering
	\caption{Detection performance of feature matrix ${F_{m \times m}}$,${P_{n \times n}}$.}
	\begin{tabular}{ccccc}
		\toprule
		\multirow{2}[4]{*}{Feature Matrix} & \multicolumn{2}{c}{Classical Dataset} & \multicolumn{2}{c}{GAUDI-DI Dataset} \\
		\cmidrule{2-5}          & Accuracy & F1-Score & Accuracy & F1-Score \\
		\midrule
		$F_{m \times m}$ & \textbf{93.07\%} & \textbf{93.73\%} & \textbf{79.55\%} & \textbf{78.33\%} \\
		$P_{n \times n}$ & 91.09\% & 91.28\% & 77.69\% & 76.15\% \\
		\bottomrule
	\end{tabular}%
	\label{tab:5}%
\end{table}%

Table \ref{tab:5} shows the experimental results of the feature matrix ${F_{m \times m}}$ on classical and GAUDI-DI. The table shows that the frequency feature matrix achieves an accuracy of 93.07\% on classical, which is significantly higher than the 83.17\% obtained for the frequency shallow feature ${F_f}$ and 92.08\% for the frequency fitting coefficient. The phase feature matrix ${P_{n \times n}}$ also has a detection of 91.09\% and 77.69\%. The frequency feature matrix is much larger than the phase feature matrix, and more information about the difference between real audio and tampered audio is obtained from the ENF frequencies through the convolutional neural network training. The detection accuracy can be further improved by fully utilizing the ENF information through the neural network.

\subsection{Validation of deep features}
This part will verify the validity of frequency and phase depth features, as shown in Fig. \ref{fig:3} (feature2, feature3). After the deep features, we perform classification by DNN classifier (two fully connected layers and one dropout layer with 1024, 256 neurons, Dropout rate=0.2, and finally one softmax layer). Meanwhile, we conducted deep phase and frequency feature fusion experiments to splice the deep features and then classify them with DNN. The experimental results are shown in the following Table \ref{tab:6}.

\begin{table}[htbp]
	\centering
	\caption{Detection performance of deep feature.}
	\begin{tabular}{lcccc}
		\toprule
		\multicolumn{1}{c}{\multirow{2}[4]{*}{Deep Feature}} & \multicolumn{2}{c}{Classical Dataset} & \multicolumn{2}{c}{GAUDI-DI Dataset} \\
		\cmidrule{2-5}          & Accuracy & F1-Score & Accuracy & F1-Score \\
		\midrule
		Frequency deep features & 94.06\% & 94.32\% & 84.46\% & 81.12\% \\
		Phase deep features & 86.14\% & 85.21\% & 78.88\% & 75.28\% \\
		Deep features fusion & \textbf{95.05\%} & \textbf{95.36\%} & \textbf{86.96\%} & \textbf{83.19\%} \\
		\bottomrule
	\end{tabular}%
	\label{tab:6}%
\end{table}%

\par
Table \ref{tab:6} shows the classification effect of deep features. It can be seen that the detection effect of frequency deep features on the two datasets is 94.06\% and 84.46\%, respectively, which is significantly better than that of frequency shallow features. The detection effect of deep phase features is comparable to that of shallow features. In addition, the accuracy of deep phase feature fusion is higher than that of single features, further demonstrating the role of feature fusion in audio tampering detection. Compared with the shallow feature ${F_{01f}}$, the deep feature fusion has higher accuracy on the GAUDI-DI dataset, indicating that the deep feature has higher generalization ability.

\begin{figure*}[ht]
  \centering
  \subfigure[Accuracy on Classical dataset]{\includegraphics[scale=0.373]{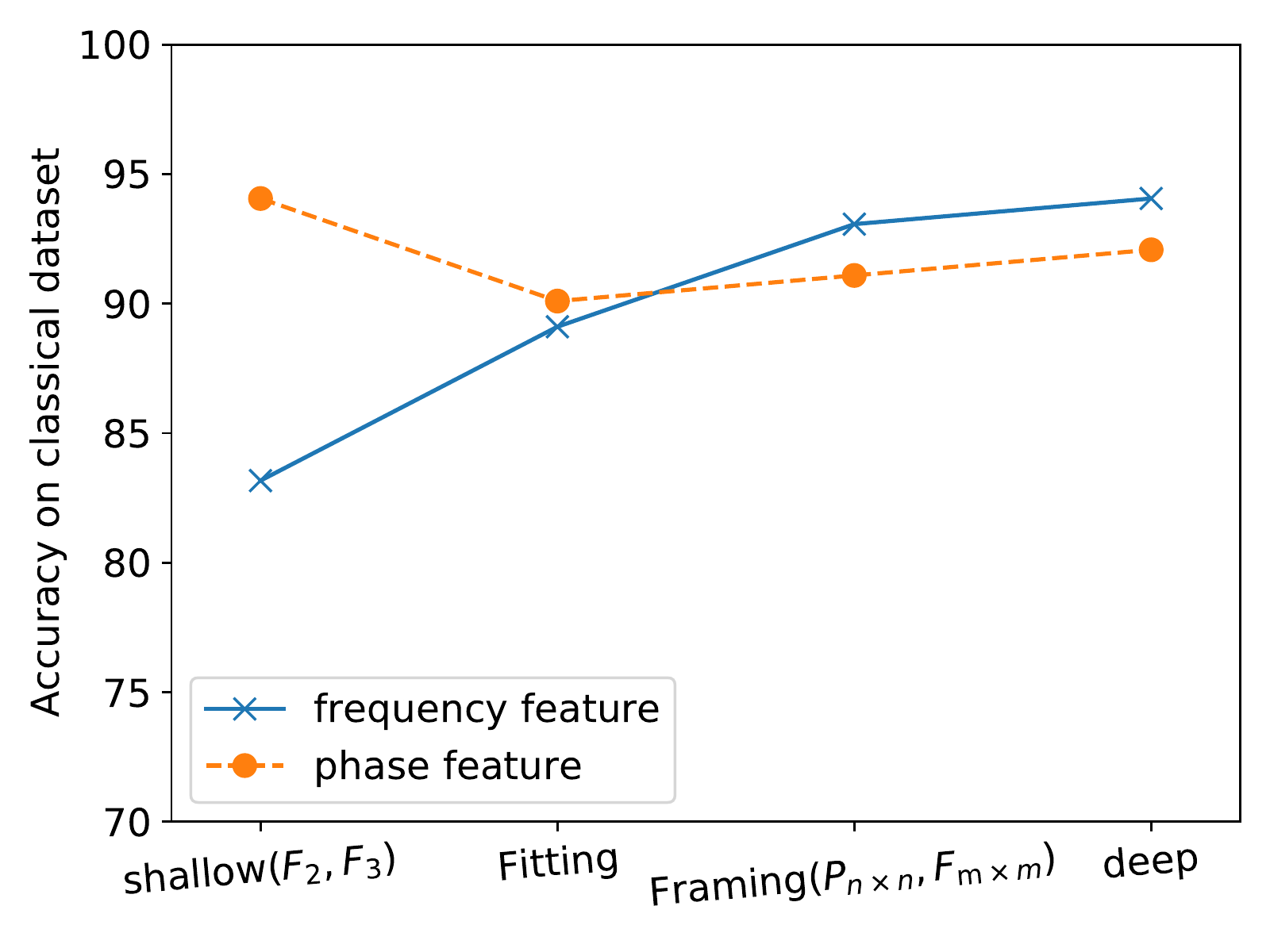}\label{fig:5a}}
  \subfigure[Accuracy on GAUDI-DI dataset]{\includegraphics[scale=0.373]{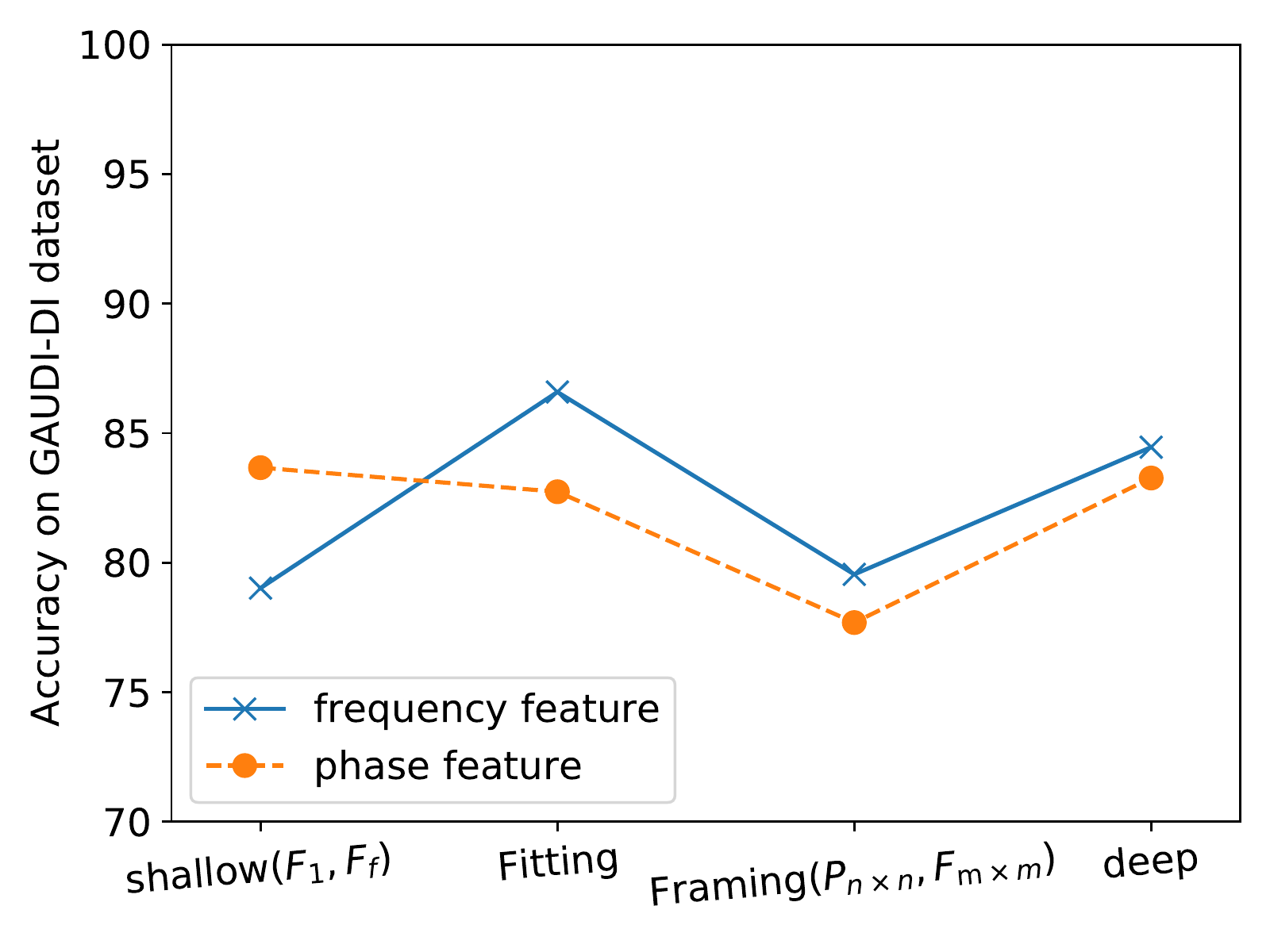}\label{fig:5b}}
  \caption{Accuracy of phase and frequency features.shallow features(${F_1}$, ${F_f}$), Fitting coefficient(${P_{coe}}$,${F_{coe}}$), Input of the convolutional(${P_{n \times n}}$, ${F_{m \times m}}$), Deep phase and frequency features}
  \label{fig:5}
\end{figure*}

\par

Also, we found that the shallow phase features outperformed the frequency features for classification, while the deep features outperformed the frequency features for classification. (As shown in Fig.\ref{fig:5}) When the detection accuracy is verified on Classical and the generalization ability is verified on GAUDI-DI, the phase ${F_1}$ of the shallow features outperforms the frequency feature ${F_f}$. The reason for the different behaviors of the two curves in Figure 4 is that the data source of Classical dataset is relatively single, while the GAUDI-DI dataset has a more complex data source, resulting in a higher accuracy on Classical dataset.

In contrast, in the deep features, the frequency feature matrix ${F_{m \times m}}$ outperforms the phase feature matrix ${P_{n \times n}}$, and the deep frequency features outperform the deep phase features. It can be judged that shallow features and deep features contain different information, and they are complementary. We use the neural network to obtain more details from the ENF, while the shallow features reflect the discontinuity information of the ENF. Therefore, we can further use the complementary characteristics of shallow and deep features to improve the model's classification accuracy and generalization ability.

\subsection{Validation of the fusion of shallow and deep features.}
In this part, the shallow and deep feature fusion methods proposed in this paper (as shown in Fig.\ref{fig:3}) are tested. The experimental variables are the final Dropout rate and the use of the Attention mechanism for feature fusion. The experimental results are shown in the following Table \ref{tab:7}.

\begin{table}[htbp]
	\centering
	\caption{Detection performance of fusion method.}
	\begin{tabular}{cccccc}
		\toprule
		\multirow{2}[4]{*}{Fusion Method} & \multicolumn{2}{c}{Classical Dataset} & \multicolumn{2}{c}{GAUDI-DI Dataset} & \multirow{2}[4]{*}{Dropout rate} \\
		\cmidrule{2-5}          & Accuracy & F1-Score & Accuracy & F1-Score &  \\
		\midrule
		\multirow{4}[2]{*}{With Attention} & 95.05\% & 95.41\% & 87.92\% & 87.71\% & 0 \\
		& 96.04\% & 96.12\% & 87.78\% & 87.46\% & 0.1 \\
		& \textbf{97.03\%} & \textbf{96.91\%} & \textbf{88.31\%} & \textbf{88.17\%} & 0.2 \\
		& 96.04\% & 96.12\% & 87.65\% & 87.32\% & 0.3 \\
		\midrule
		\multirow{4}[2]{*}{Without Attention} & 94.06\% & 94.33\% & 87.38\% & 87.13\% & 0 \\
		& 95.05\% & 95.41\% & 88.05\% & 87.82\% & 0.1 \\
		& \textbf{95.05\%} & \textbf{95.41\%} & \textbf{87.92\%} & \textbf{87.75\%} & 0.2 \\
		& 95.05\% & 95.41\% & 87.25\% & 87.06\% & 0.3 \\
		\bottomrule
	\end{tabular}%
	\label{tab:7}%
\end{table}%

\par
Table \ref{tab:7} shows the experimental results of this paper with feature fusion as the experimental variable. The table shows that the experimental results on two different datasets show that the proposed method achieves the highest detection accuracy and model generalization ability at dropout rate=0.2. The feature fusion with attention is better than the mechanism without attention. The model generalization ability of the proposed method is significantly better than that of the method in \cite{Wang2018}. The fusion of shallow and deep level features by the attention mechanism allows the full exploitation of the ENF phase and instantaneous frequency features. The complementary nature of the features at different levels is exploited to more accurately characterize the changes in inconsistency produced by tampering operations to natural digital audio. The attention fusion mechanism in this paper can automatically learn can give different weights to each feature value of shallow and deep features. The features significantly impacting the classification result will be given a larger weight. In comparison, the features that have an insignificant effect on the final classification will be given a smaller weight to improve the detection accuracy and generalization ability.

\subsection{Discussion}
In this section, we conducted five sets of experiments. In experiment 1 (comparison of the proposed method with the state-of-the-art methods), we concluded that the proposed method is better than the state-of-the-art method in \cite{Wang2018}, and the model generalization ability is significantly better than the baseline methods; in experiment 2 (validation of the fitting coefficient features), we concluded that for the duration of the audio to be measured is 9-35s, the global information compensation as the deep features; in experiment 3 (validation of feature matrix ${F_{m \times m}}$,${P_{n \times n}}$) and experiment 4 (validation of deep features), we verify the validity of the deep features proposed in this paper; in experiment 5 (validation of the fusion of shallow and deep features), we verify the effectiveness of feature selection by attention mechanism in this paper.

The results show that :1). Through the fusion of ENF phase and frequency features, audio tampering detection can achieve higher detection accuracy by using different information contained in the ENF phase and frequency in audio. 2). The shallow features contain ENF discontinuity information, while the deep features obtained by the deep learning method contain more ENF details. The complementary feature of the shallow features and the deep features can make the tampering detection method have higher accuracy and generalization ability. 3). The attention mechanism was used for feature fusion, and different weights were assigned to each feature value to suppress invalid features, which further improved the model's performance.

\section{Conclusion} \label{conclusion}
This paper proposes an audio tampering detection method based on the fusion of shallow and deep features. Firstly, the phase and instantaneous frequency characteristics of ENF in audio were obtained by DFT and Hilbert transform. Then we obtained the shallow layer characteristics reflecting ENF discontinuity through calculation and obtained the deep phase and frequency characteristics through the neural network. Finally, the attention mechanism is used for feature fusion. After dimensionality reduction, the Softmax classifier is used for classification to detect the edited audio. By fusing shallow and deep features, the complementarity of features at different levels is exploited to more accurately describe the changes in inconsistency produced by tampering operations to raw digital audio. Further, attention is used to fuse phase features and frequency features to obtain rich information from audio ENF for tampering detection classification tasks through automatic learning to improve detection accuracy and model generalization. Experimental results show that the proposed method has higher recognition accuracy and generalization ability. Future work will focus on more robust audio tampering detection methods. In addition, detection methods will be designed to locate the location of the audio tamper.

\begin{backmatter}

\section*{Abbreviations}
ENF: electronic network frequency; STFT: short time Fourier transform; MFCCs: Mel-frequency cepstral coefficients; DFT: discrete Fourier transform; TPSW: two-pass Split Window; CNN: convolutional neural network; SVM: support vector machine 

\section*{Funding}
The research work of this paper were supported by the National Natural Science Foundation of China (No. 62177022, 61901165, 61501199), Collaborative Innovation Center for Informatization and Balanced Development of K-12 Education by MOE and Hubei Province (No. xtzd2021-005), and Self-determined Research Funds of CCNU from the Colleges’ Basic Research and Operation of MOE (No. CCNU20ZT010).

\section*{Availability of data and materials}
Please contact authors for data requests.

\section*{Competing interests}
The authors declare that they have no competing interests.

\section*{Authors' contributions}
Equal contribution from all authors. All authors read and approved the final manuscript.

\section*{Acknowledgements}
The authors acknowledge the comments by anonymous reviewers that helped to improve a preliminary version of the paper.


\bibliographystyle{bmc-mathphys} 
\bibliography{bmc_article}      

\end{backmatter}
\end{document}